\documentclass[prb,showpacs,twocolumn,preprintnumbers,amsmath,amssymb,floatfix]{revtex4}
\usepackage{graphicx}
\usepackage{dcolumn}
\usepackage{bm}
\usepackage{graphics}
\usepackage{amssymb}
\usepackage{amsmath}
\usepackage{xspace}
\usepackage{epsfig}
\usepackage{longtable}
\newcommand {\etal}{\begin{itshape}et al\end{itshape}. }

\newcommand {\ibid}{\begin{itshape}ibid\end{itshape}., }

\renewcommand {\vec}{\mathbf}

\newcommand {\lY}{$\lambda$-BETS$_2$Y }
\newcommand {\kX}{$\kappa$-ET$_2$X }

\newcommand {\bIBr}{$\beta$-ET$_2$IBr$_2$ }

\newcommand {\Br}{$\kappa$-ET$_2$Cu[N(CN)$_2$]Br }

\newcommand {\NCS}{$\kappa$-ET$_2$Cu(NCS)$_2$ }

\begin{document}
\title{On the Relationship Between the Critical Temperature and the London Penetration Depth in Layered Organic
Superconductors}
\author{B. J. Powell}
\email{powell@physic.uq.edu.au}
\affiliation{Department of Physics, University of Queensland,
Brisbane, Queensland 4072, Australia} 
\author{Ross H. McKenzie}
\affiliation{Department of Physics, University of Queensland,
Brisbane, Queensland 4072, Australia} 

\begin{abstract}
We present an analysis of previously published measurements of the
London penetration depth of layered organic superconductors. The
predictions of the BCS theory of superconductivity are shown to
disagree with the measured zero temperature, in plane, London
penetration depth by up to two orders of magnitude. We find that
fluctuations in the phase of the superconducting order parameter
do not determine the superconducting critical temperature as the
critical temperature predicted for a Kosterlitz--Thouless
transition is more than an order of magnitude greater than is
found experimentally for some materials. This places constraints on theories of
superconductivity in these materials.
\end{abstract}

\maketitle

In this paper we consider the layered organic superconductors such
as \kX and \lY (Ref. \onlinecite{footnote:ET_BETS}). Most theories
of superconductivity in these materials are based on BCS theory
with either phonons \cite{Phonons} or spin fluctuations
\cite{SF_BCS,SF_inst} providing the attractive interaction.
However, we will show that simple BCS theory is inconsistent with
the measured London penetration depth.
\cite{Pratt_nU,Lang_penetration2,Larkin_penetration_depth} Layered
organic superconductors are, in many ways, similar to the
cuprates. \cite{Ross_science} Both classes of materials are
quasi-two dimensional (q2D) and have phase diagrams which include
antiferromagnetism, a Mott transition, unconventional metallic
states and superconductivity. The superconducting state of the
cuprates has d-wave symmetry \cite{Tsuei} and, although there is,
as yet, no consensus \cite{BenRHMdisorder} on the pairing symmetry
in the organics, several authors have presented
evidence for d-wave pairing. \cite{BenRHMdisorder} 
NMR experiments on the layered organic superconductors are
suggestive of a pseudogap \cite{NMR} similar to that observed in
the cuprates. \cite{Loktev} It has been suggested that the Hubbard
model is a minimal model for both of these systems.
\cite{Ross_review,Anderson_RVB} The most notable difference
between the two classes of materials is that in the cuprates
doping changes the charge carrier density. Whereas the organics
are, as we will confirm, half filled for all the anions that we
consider here.

It has been suggested \cite{EK} that fluctuations in the phase of
the superconducting order parameter determine the superconducting
critical temperature, $T_c$, in both the (underdoped) cuprates and
the layered organic superconductors. In this paper we will show
that recent experimental data
\cite{Pratt_nU,Lang_penetration2,Larkin_penetration_depth}
disproves this conjecture in the case of the layered organic
superconductors and discuss which theories are consistent with
these experiments.

It is widely believed that the anion layers of the layered organic
superconductors are insulating and that at low temperatures the
metallic phase of the organic layers can be described by a Fermi
liquid tight-binding model that is half-filled.\cite{Ross_review}
A check of this model is to compare its predictions with the size
of the orbits observed in quantum oscillation experiments. The
area, $A$, enclosed by an orbit in a quantum oscillation
experiment is related to the observed frequency, $F$ in $1/B$,
where $B$ is the magnetic field strength, by the Onsager
relationship, $A=\frac{2\pi e}{\hbar}F$. Thus for a q2D Fermi
liquid it follows from Luttinger's theorem that the electron
density $n_e=\frac{\hbar}{4\pi^3e}F$. 

The q2D area occupied by a dimer, $A_d$, can be calculated from
crystallographic measurements (see the caption to table
\ref{table}). If one assumes that each dimer donates exactly one
electron to each anion then the product $n_eA_d$ is predicted to
be unity. It can be seen from table \ref{table} that this
prediction is in excellent agreement with experiment.

\begin{table*}
\caption{\label{table} The electron and superfluid densities of
various layered organic superconductors. $m_\beta^*/m_e$ is the
effective mass of the magnetic breakdown ($\beta$) orbit
determined from quantum oscillation experiments. $n_e$ is the
electron density calculated via the Onsager relationship
from quantum oscillation experiments. The quasi-two dimensional
area occupied by a dimer is $A_d=V_{uc}/N_dd$, where $V_{uc}$ is
the volume of the unit cell, $N_d$ is the number of dimers per
unit cell and $d$ is the average interlayer spacing. Thus for a
quasi-two dimensional tight binding model of dimers at half
filling one expects $n_eA_d=1$. This is indeed observed
experimentally. This shows that there is no correlation between
the band filling and the many-body effects responsible for the
mass renormalisation. We have taken both $T_c$ and $\lambda_0$
from the same experiments as both quantities can be sample
dependent. \cite{BenRHMdisorder}
The superfluid density defined by $n_s = m^*c^2/4\pi
e^2\lambda_0^2$. Note that $n_s/n_e$ varies approximately linearly
with $T_c$ and that the BCS prediction that $n_s=n_e$ is strongly
violated by the low $T_c$ materials: $\beta$-(ET)$_2$IBr$_2$,
$\alpha$-(ET)$_2$NH$_4$Hg(NCS)$_4$ and
$\kappa$-(BETS)$_2$GaCl$_4$.}
\begin{center}
\begin{tabular}{lcccccccc}
\vspace*{-0.65cm}
\\\hline \hline \vspace*{-8pt}  \\
Material&$n_{e}$ (nm$^{-2}$) \hspace*{2pt} & \hspace*{2pt} $A_d$
(nm$^{2}$)\hspace*{2pt} & \hspace*{2pt} $n_eA_d$ \hspace*{2pt} &
\hspace*{2pt} $T_c$ (K) \hspace*{2pt} & \hspace*{2pt} $\lambda_0$
($\mu$m) \hspace*{2pt} & \hspace*{2pt} $m_\beta^*/m_e$
\hspace*{2pt} & \hspace*{2pt} $n_s$ (nm$^{-2}$) \hspace*{2pt} &
\hspace*{2pt} $n_s/n_{e}$ \\
\hline \vspace*{-9pt} \\
$\kappa$-(ET)$_2$Cu[N(CN)$_2$]Br & 1.83\cite{Mielke_Br} &
0.552\cite{Geiser} & 1.01 & 11.6\cite{Le1} & 0.78\cite{Le1} &
6.4\cite{Weiss} & 0.64 & 0.35\\
$\kappa$-(ET)$_2$Cu(NCS)$_2$ & 1.83\cite{Caulfield} &
0.519\cite{Urayama} & 0.95  & 9.4\cite{Pratt_nU} &
0.54\cite{Pratt_nU} & 6.5\cite{Caulfield} & 1.0 & 0.55\\
$\lambda$-(BETS)$_2$GaCl$_4$ & 1.95\cite{Mielke} &
0.484\cite{Tanaka} & 0.94  & 5.5\cite{Pratt_nU} &
0.72\cite{Pratt_nU} & 6.3\cite{Mielke} & 0.63
& 0.32\\
$\beta$-(ET)$_2$IBr$_2$ & 1.91\cite{Wosnitza_ratio_kIBr} &
0.549\cite{Williams} & 1.05  & 2.21\cite{Pratt_nU} &
0.90\cite{Pratt_nU} & 4.0\cite{Wosnitza_ratio_kIBr}
& 0.21 & 0.11\\
$\alpha$-(ET)$_2$NH$_4$Hg(NCS)$_4$ & 1.95\cite{Polisskii}  &
0.488\cite{Mori} & 0.95  & 1.12\cite{Pratt_nU} &
1.09\cite{Pratt_nU} & 2.0\cite{Polisskii} & 0.098 & 0.050\\
$\kappa$-(BETS)$_2$GaCl$_4$ &  2.11\cite{Tajima} & ? & ? &
0.16\cite{Pratt_nU} & 2.26\cite{Pratt_nU} &
2.4\cite{Tajima} & 0.025 & 0.012\\
\hline\hline \vspace*{-1cm}
\end{tabular}
\end{center}
\end{table*}

In general, a state is deemed superconducting if it breaks gauge
symmetry and displays a Meissner effect in weak magnetic fields.
It follows directly from these very general requirements that a
supercurrent, $\vec{j}=-D_s2e\vec{A}/\hbar\equiv
-c^2\vec{A}/4\pi\lambda^2$, is induced by a magnetic vector
potential $\vec{A}$. $\lambda$ is the London penetration depth and
$D_s$ is the superfluid stiffness. In BCS theory and its
extensions one can separate $D_s$ into a superfluid density and an
effective mass, ($D_s\propto n_s/m^*$). Here $m^*$ is the
effective mass of the quasiparticle excitations and $n_s$
describes the proportion of electrons in the condensate in the
terms of the two-fluid model. However, this separation is
\textit{not} a necessary feature of a superconducting state.
\cite{Scalapino}

In London theory the zero temperature superfluid density 
is defined as \cite{Tinkham} $n_s = m^*c^2/4\pi e^2\lambda_0^2$,
where $\lambda_0$ is the average London penetration depth parallel
to the q2D planes at zero temperature. BCS theory
\cite{Tinkham} 
 predicts that $n_s=n_e$ and
Eliashberg theory \cite{Blezius} predicts that $n_s\lesssim n_e$.
It can be shown \cite{Millis} that, for a charged system,
including the Fermi liquid corrections to BCS theory gives
\begin{eqnarray}
\frac{n_s}{n_e}=\frac{1+\frac{1}{3}F_1^s}{m^*/m},
\label{eqn:renorm_ns}
\end{eqnarray}
where $F_1^s$ is a Landau Fermi liquid parameter. For a Galilean
invariant system $1+F_1^s/3=m^*/m$ and so $n_s=n_e$. But for
systems with broken translational symmetry, such as the crystals
that we consider here, there is no \emph{a priori}
relationship\cite{Millis} between $F_1^s$ and $m^*$.

It can be seen from table \ref{table} that the predictions of BCS
theory are in disagreement with experiments on the layered organic
superconductors by up to two orders of magnitude. It has been
suggested that only the q2D pocket of the Fermi surface of \Br is
involved in superconductivity. \cite{Mielke_Br} Such Fermi surface
sheet dependent superconductivity can be ruled out as the
explanation of the reduced superfluid density because, for
example, the Fermi surface of \bIBr ($n_s/n_e=0.11$) has only one
sheet. \cite{Wosnitza_ratio_kIBr} Corrections due to the variation
in the Fermi velocity around the Fermi surface, \cite{vF} may be
able to explain small deviations from $n_s/n_e=1$, but are
certainly not large enough to explain the extremely small
superfluid density observed in the low $T_c$ materials.

The simplest explanation of the penetration depth measurements is
that not all of the electrons participate in the superconducting
condensate. This would lead to many observable predictions. For
example, thermodynamic indications of the superconducting state
would be expected to show a \lq mixed' behaviour, e.g. the
specific heat anomaly and the effective Meissner volume should be
extremely small in low $T_c$, low $n_s$ compounds. Thus the
observation of a clear anomaly in the heat capacity \cite{Cv} and
a complete Meissner effect \cite{Taniguchi} in
$\alpha$-(ET)$_2$NH$_4$Hg(NCS)$_4$ ($n_s/n_e=0.05$) appear to rule out scenarios in
which only a fraction of the conduction electrons enter the
condensate. Another possibility that retains the independent
concepts of the effective mass and the superfluid density is to
allow the Cooper pair to have an effective mass that is not simply
$2m^*$. This has been discussed elsewhere and we will not dwell on
this idea here as it was shown \cite{Pratt_ISCOM} that even in
these scenarios it is still necessary to set $n_s/n_e\ne1$ to
explain the observed behaviour of the layered organic
superconductors.

Note that for the organics the superfluid density is smallest for
those materials with the lowest $T_c$'s and the smallest effective
masses, i.e., those materials that are the least strongly
correlated. This is in direct contradiction with the predictions
of the simple interpretations of the BCS and
Eliashberg\cite{Blezius} theories where as the electron-phonon (or
indeed electron-electron) coupling increase so do $m^*$ and $T_c$.
In the underdoped cuprates the pseudogap is associated with low
critical temperatures and small superfluid densities, whereas in
the organics the pseudogap like features are associated with high
critical temperatures and large superfluid densities. However, in
both classes of materials the pseudogap is found close to the Mott
transition.


It has been suggested\cite{Millis} that in the cuprates $F_1^s$
increases as $m^*$ increases, rather than in the decreasing as is
the case for a Galilean invariant Fermi liquid. Could a similar,
albeit significantly stronger, effect be at play here? If
$n_s\rightarrow0$ as $T_c\rightarrow0$ (while at the same time
$m^*$ decreases) then (\ref{eqn:renorm_ns}) requires that
$F_1^s\rightarrow-3$ as $T_c\rightarrow0$. For a momentum
independent self energy $F_1^s=0$ (Ref. \onlinecite{Engelbrecht});
therefore for either the BCS or Eliashberg theories to be
consistent with the data would require a strong momentum
dependence in the self energy. Electron phonon coupling can only
generate a momentum dependent self energy if Migdal's theorem is
strongly violated.\footnote{Note that none of the phonon mechanism
that have been discussed in the context of the
organics\cite{Phonons} have proposed that Migdal's theorem is
broken.} 
However, a strong momentum
dependent self energy may be a more natural feature of spin
fluctuation mediated superconductivity.\cite{SF_BCS,SF_inst}

In the case of very strong electron-electron interactions equation
(\ref{eqn:renorm_ns}) may not be valid. However, importantly,
unlike underdoped cuprates, in the organics the normal state at
temperatures only slightly above $T_c$ appears to be a good Fermi
liquid.\cite{Ross_review} 
Hence, there is a
need to calculate $D_s$ for the models and approximations that
have been proposed for the organic
superconductors\cite{Phonons,SF_BCS,SF_inst} to see is they
predict the observed variation in $D_s$ with $T_c$.

A possible explanation of the measured penetration depths is that
the microscopic theory of superconductivity in the layered
organics, whatever it may be, does not admit the separation of the
superfluid stiffness into parts that correspond naturally to a
superfluid density and an effective mass. This has the advantage
of allowing the observation of a small superfluid stiffness to be
reconciled with evidence that all of the electrons participate in
the condensate.

To explain the Uemura relation, \cite{Uemura91} namely that in the
underdoped cuprates $T_c\propto1/\lambda_0^2$, Emery and Kivelson
\cite{EK} proposed that phase fluctuations can limit the
transition temperature of a q2D superconductor. The limit on $T_c$
due to phase fluctuations, $T_\theta^\textrm{max}$, is given by
\begin{eqnarray}
k_BT_\theta^\textrm{max} = A\frac{\hbar^2c^2a}{16\pi
e^2\lambda_0^2}, \label{eqn:EK}
\end{eqnarray}
where $a$ is the larger of $d$, the average spacing between the
q2D planes, and $\sqrt{\pi}\xi_\perp$, where $\xi_\perp$ is the
coherence length perpendicular to the planes. $A$ is a constant of
order 1. In the case of vanishingly small coupling between the
planes we have a genuinely two dimensional system and therefore
the superconducting transition is a Kosterlitz--Thouless phase
transition. In the underdoped cuprates further support for these
ideas comes from measurements of the optical conductivity
\cite{Corson} of Bi$_2$Sr$_2$CaCu$_2$O$_{8+\delta}$ for $T>T_c$
which are consistent with the predictions of Kosterlitz--Thouless
theory and the observation of vortex like excitations \cite{Xu}
above $T_c$ in La$_{2-x}$Sr$_x$CuO$_4$. However, we should note
that the evidence of phase fluctuations in these experiments did
not extend to temperatures as high as those at which the onset of
the pseudogap is observed. \cite{Carlson_review}

\begin{figure}
    \centering
    \epsfig{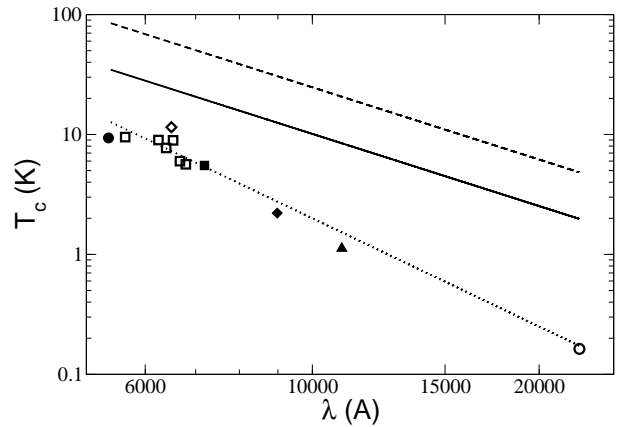}
    \caption{Variation of the superconducting critical
    temperature, $T_c$,
    with the zero temperature penetration depth, $\lambda_0$.
    The experimental data is taken from
    Pratt \textit{et al}., \cite{Pratt_nU} Lang \textit{et al}. \cite{Lang_penetration2}
    and Larkin \textit{et al}. \cite{Larkin_penetration_depth} and shows data for
    $\kappa$-ET$_2$Cu[N(CN)$_2$]Br (open diamond)\cite{Lang_penetration2}
    $\kappa$-ET$_2$Cu(NCS)$_2$ both at ambient pressure (circle)\cite{Pratt_nU}
    and under pressure (open squares),\cite{Larkin_penetration_depth} $\lambda$-BETS$_2$GaCl$_4$
    (square), $\beta$-ET$_2$IBr$_2$ (diamond),\cite{Pratt_nU}
    $\alpha$-ET$_2$NH$_4$Hg(NCS)$_4$ (triangle)\cite{Pratt_nU} and
    $\kappa$-BETS$_2$GaCl$_4$ (empty circle). \cite{Pratt_nU}  The empirical fit,
    $T_c\lambda^3=2.0$~K$\mu$m$^3$, to the
    data from Pratt \textit{et al}. is also reproduced (dotted dashed line).
    Note that the
    data of Larkin \textit{et al}. (open squares) is actually for the penetration
    depth at $T=0.35T_c$. This means the data should be shifted
    somewhat to the left. However, even given this caveat the pressure dependence data of Larkin \textit{et al}.
    is in broad agreement with the ambient pressure data of Pratt
    \textit{et al}.
    The upper limit imposed on $T_c$ by
    phase fluctuations, $T_\theta^\textrm{max}$, (\ref{eqn:EK}) is shown for both the three dimensional
    (dashed line, $A=2.2$) and two dimensional (solid
    line, $A=0.9$) cases. Although it is possible that the details of the short-range interactions of the
    layered organic superconductors change the exact numerical values of $A$ (c.f., Ref.
    \onlinecite{Carlson}),
    it is difficult to imagine that this effect is large enough to account for the order
    of magnitude difference between the predictions of the phase fluctuation model
    and the observed variation of  $T_c$
    with $\lambda_0$. For the phase fluctuation curves (solid and dashed lines) we take
    $a=d=18$~\AA, where $a$ is the length parameter in equation (\ref{eqn:EK})
    and $d$ is the interlayer spacing which approximately
    18~\AA~ for all of these materials.} \label{fig:Tc_lambda}
\vspace*{-15pt}
\end{figure}

Emery and Kivelson \cite{EK} suggested that the data of Uemura
\etal \cite{Uemura91} implies that the critical temperatures of
the layered organic superconductors in general and of \NCS in
particular are also limited by phase fluctuations. In figure
\ref{fig:Tc_lambda} we plot $T_c$ as a function of $\lambda_0$ for
a variety of layered organic superconductors. It can clearly be
seen that $T_\theta^\textrm{max}$ is more than an order of
magnitude larger than $T_c$ for some of the materials considered
(c.f., Ref. \onlinecite{Pratt_ISCOM}). 

Further evidence that $T_c$ is limited by the temperature at which
pairing occurs and not by the energy scale of phase fluctuations
comes from the ratio of the zero temperature superconducting order
parameter, $\Delta(0)$, to $T_c$. For \Br and \NCS it has been
found that \cite{alpha} that $\Delta(0)/k_BT_c=2.5-2.8$. These
values of $\Delta(0)/k_BT_c$ seem more consistent with strong
coupling superconductivity than with the expectation
\cite{Kivelson_private} that, if $T_c$ is limited by phase
fluctuations, $\Delta(0)/k_BT_c\gg2$, which is indeed found for
the underdoped cuprates. Measurements of $\Delta(0)/k_BT_c$ in low
$T_c$ materials may be expected provide a more stringent test of
this criterion, however, we are not aware of any such
measurements.

The destruction of superconductivity by phase fluctuations is
strongly linked with the idea that preformed pairs are responsible
for the pseudogap in the cuprates. \cite{Corson,Xu,EK,Loktev}
Therefore the observation that $T_c$ is not limited by phase
fluctuations in the layered organics makes it unlikely that
preformed pairs are responsible for the pseudogap like features
observed by NMR. \cite{NMR}

For the cuprates several theories have been proposed that may
admit an increase in the superfluid stiffness as one moves away
from the Mott insulating phase by increasing the doping from half
filling. Examples of these include the RVB state
\cite{Anderson_RVB} and its generalisation gossamer
superconductivity, \cite{gossamer} the SU(2) slave-Boson model
\cite{Wen&Lee} and the two-species treatment of the $t$-$J$ model.
\cite{Baskaran} Thus the observation that the superfluid stiffness
varies as one moves away from the Mott insulating phase in the
layered organic superconductors may indicate that one of these
theories provides the correct microscopic description of these
materials. 
Clearly detailed calculations are required to discover whether any of 
these models agree with the experimentally measured penetration depth.

It appears then that the key to understanding the microscopic
details of the superconducting state in the organic
superconductors is the low $T_c$ materials. In addition to the
need for a detailed systematic, study of the thermodynamics of the
low $T_c$ materials discussed here there are several other
materials with low abient pressure $T_c$s that should be
investigated such as $\beta$-ET$_2$AuI$_2$ ($T_c=4.9$~K),
$\kappa$-ET$_2$I$_3$ ($T_c=3.6$~K), $\lambda$-BETS$_2$GaCl$_3$F
($T_c=3.5$~K), $\kappa$-DMET$_2$AuBr$_2$ ($T_c=1.9$~K),
BO$_2$Re$_4\cdot$H$_2$O ($T_c=1.5$~K) and
$\beta$-BO$_3$Cu(NCS)$_3$ ($T_c=1.1$~K).

We have shown that the zero temperature superfluid stiffness of
the layered organic superconductors is up to two orders of
magnitude smaller than is predicted by simple BCS theory. We have
also shown that phase fluctuations do not limit $T_c$ in these
materials as the transition temperature is more than an order of
magnitude smaller than is predicted for a Kosterlitz--Thouless
phase transition. 
This places constraints on 
theories of superconductivity in layered organic
superconductors. It is therefore clear that the unusual behaviour
of the penetration depth is a key experimental result which any
theory of the layered organic superconductors must explain.

This work was stimulated by discussions with Francis Pratt. We
would like to thank Tony Carrington, G. Baskaran and Steven
Kivelson for illuminating discussions. This work was supported by
the Australian Research Council.


\begin{thebibliography}{72}
\bibitem{footnote:ET_BETS} ET is bis(ethylenedithio)tetrathiafulvalene, BETS is
bis(ethylenedithio)tetraselenafulvalene, X and Y are anions, e.g,
X=Cu(NCS)$_2$ or Y=GaCl$_4$ and the Greek indices indicate crystal
structure; see Refs. \onlinecite{Ross_review} for a recent review.
\end{thebibliography}
\end{document}